\documentclass[11pt,a4paper]{article}
\usepackage[nohead, nomarginpar, margin=1in, foot=.25in]{geometry}

\usepackage{graphicx}
\usepackage{color}

\usepackage{amsmath}
\usepackage{amsmath,bbm}

\usepackage{amssymb,bm}

\usepackage{comment}

\usepackage[hyphens]{url} 

\usepackage{siunitx} 
\usepackage{csquotes} 

\usepackage{subcaption}

\usepackage{multirow}
\usepackage{tabularx} 
\usepackage{tabulary} 

\usepackage[T1]{fontenc}
\usepackage[utf8]{inputenc}
\usepackage{authblk}

\usepackage{soul}
\usepackage[LGRgreek]{mathastext}

\usepackage{url}

\providecommand{\keywords}[1]{\textit{Key words---} #1}
\providecommand{\pacs}[1]{\textit{JEL codes---} #1}

\usepackage{cite}
\date{}

\def\@listcomma@comma{\@ifnum{\@tempcnta>\tw@}{,}{}}

\begin{document}

\title{Impact of crop diversification on socio-economic life of tribal farmers: A case study from Eastern ghats of India}
\author{Sadasiba Tripathy\footnote{tripathy.sadasiba@gmail.com}}
\author{Dr. Sandhyarani Das}
\affil{Berhampur University, Berhampur-760007, Odisha, India}

\maketitle

\begin{abstract}
In this study we investigated impact of crop diversification on socio-economic life of tribal people from eastern ghats of India. We have adopted linear regression formalism to check impact of cross diversification. We observe a positive intercept for almost all factors. Coefficient of correlation is calculated to examine the inter dependence of CDI and our various individually measured dependent variables. A positive correlation is observed in almost all factors. This study shows that a positive change occurred in their social, economic life in the post diversification era.
\end{abstract}

\noindent
 \\
 \\

\section{Introduction}
\label{intro}

\subsection{Socio-economic life and tribal farming}
\label{Cropdiversificationandtribal farming}
Agriculture in tribal life is not a practice rather a culture related to their social set up and traditional beliefs. They believe the production of the crop is the gift of nature. So they worship the land and observe the festivals related to the crop harvesting in different seasons \cite{tribe_AgriIsFestival}.
Also they worship nature to maintain a productive climatic condition during the period of cropping and harvesting \cite{tribe_worshipAtstartAgri}. 
Day by day the farmers are getting conscious about the market situations may be due to the demonstration affect of the non-tribal farmers or from different Govt agencies \cite{tribe_govtAgenciesRole}. They have now diversified their cropping pattern from paddy to non-paddy. They realized that the diversification of crops from traditional crops to vegetable cropping lead to increase in income \cite{cdi_mypaper} and also realized the factors like irrigation , fertilizer , rainfall , farmers knowledge , etc are highly responsible for crop diversification in the study area.  Diversification is now became an assert of economic growth in the tribal areas of Odisha. Though there are instances of uneven rainfall leads to crop loss and crop concentration but they should go on diversifying their cropping pattern to continue with the average growth of economy and per capita in the long run \cite{climatechange_mypaper}.

The increase in income leads to raise the effective demand. Here we can add that the farmers are adding to their consumption habit. They now adopting the use of the needs of modern life. In agricultural practices they started using capital intensive techniques. The changed income out of crop diversification also affects their economics and cultural life. This changing practice of spending will positively boost the economy.

To investigate this effect, we have chosen Eastern ghats of India's have tribal dominated areas. 
We have taken purposive sampling and chosen 5 different tribal dominated villages from 5 different blocks of Koraput District (18.8135\si{\degree} N, 82.7123\si{\degree} E), (Raniput/Village 1, Soralguda/Village2, Mundaguda/Village 3, Ekomba/Village 4 and Rupabeda/Village 5). The data was collected from 422 numbers of households i.e. 80\% the total households of the villages through simple random sample method. The samples were interviewed in order to deduce the inferences about their socio economic life in the post crop diversification era. 

The paper aims at analyzing the impact of crop diversification on the socio economic life of the tribal people in the study area and to know the socio economic life of the inhabitants of the land.

\subsection{Profile of the Study Area}
\label{ProfileoftheStudyArea}

The data which is represented over here is the census data of 2011 \cite{CensusofIndia2011}. These are very small villages and the number of households are 134 is the highest and 68 is the lowest. The average literacy of these 5 villages is 35.15\% which is a matter of concern. The average population of these 5 villages are 439 and average family members per household are 4.16. The range of the schedule caste people from 0 to 21.04 and schedule tribe is from 64.77 to 99.56 percent. The average of child population per village is 86.6.

\begin{table}[h!]
\centering
\caption{Demographic profile as per 2011 census \cite{CensusofIndia2011,cdi_mypaper}}
\begin{tabular}{ c|c|c|c|c|c}  
   \hline
Particulars &Village 1 &Village 2 &Village 3 &Village 4 &Village 5 \\
   \hline
  No. of families &103 &122 &134 &100 &68 \\
     \hline
Total family members &495 &423 &608 &442 &227 \\
     \hline
Child (0-6 age) &108 &84 &130 &71 &41 \\
     \hline
     Schedule Tribe (in \%) &65.05 &64.77 &77.3 &80.77 &99.56 \\
     \hline
Schedule Caste (in \%) &20.08 &21.04 &0. &0.97 &0. \\
     \hline
Literacy (in \%) &38.76 &36.58 &35.36 &53.91 &16.13 \\
     \hline
\end{tabular}
\label{Tab:VillageCensus2011}
\end{table}


Here in table -2 we can see there are two different type of working population and named them as main and marginal workers. The main workers have engaged themselves more than six months a year but the marginal workers only have work less than 6 months. The data speaks the average number of the main workers in these 5 villages is 237.4 out of which 46.43\% are the main workers and the 53.57\% are the marginal workers.

\begin{table}[h!]
\centering
\caption{Work profile as per 2011 census \cite{CensusofIndia2011,cdi_mypaper}}
\begin{tabular}{ c|c|c|c|c|c}  
   \hline
Particulars &Village 1 &Village 2 &Village 3 &Village 4 &Village 5 \\
   \hline
  Total Workers &215 &244 &317 &267 &144\\
     \hline
Main Worker (in \%) &45.12 &47.95 &47 &35.12 &56.94\\
     \hline
Marginal Worker (in \%) &54.88 &52.05 &53 &64.79 &43.06\\
     \hline
\end{tabular}
\label{Tab:WorkProfile2011}
\end{table}

\section{Model}
\label{Methodology}
Here we have taken crop diversification as the independent variable and farmer's income as dependent variable. Here, we would analyze change in the crop diversification leads to the change in farmer's income.\\
Y = $b_0+b_1X_1+\epsilon$ \cite{linearregessioneqn}\\
Where, Y = income (dependent variable),\\
$b_0$ = estimate of the regression intercept,\\
$b_1$ = estimate of the regression slope,\\ 
$X_1$ = crop diversification (Independent variable),\\
and $\epsilon$ is error term

As have few factors that depicts socio-economic life of tribal farmers, so each dependent variables are dealt by SLR individually. We calculated correlation between these coefficient by Karl Pearsons method \cite{KarlPearsonMethod}. One can get coefficient of correlation is ratio between covariance of two variables with product of their standard deviation (direct method).
\begin{equation}
r = \frac{\sum{xy}}{N \sigma_x \sigma_y}
\end{equation}
Where, x = X - $\bar{X}$ and  $\bar{X}$ is mean of X,\\
$\sigma_x$ = $\sqrt{\frac{\sum x^2}{N}}$,\\
and same for y.\\
Value of r equal to zero refers no correlation is there, and 1 represents perfect correlation.

\section{Explanation of Dependent variables}
\label{Results_and_discussion}

\subsection{Agricultural practices:}
In the pre diversification era 2004-05, the farmers used to apply the bio fertilizer and soil of the forest with cow dung in their corn fields to make the land more fertile as shown Tab. \ref{Tab:AgriculturePrctices}. They use the different leafs and hay and ash and prepare a compost to use as the fertilizer \cite{tribal_DontUseChemFert}. They collect the soil from the forests and use it in the land to make it more fertile.
 Here in Tab. \ref{Tab:AgriculturePrctices} we have taken only the cow dung manure because the bio fertilizer and soil of the forests are used in a lesser quantity. Day by day the maintenance of the cattle is getting very complex and people are adopting the modern day agricultural practices and adopted the use of chemical fertilizers in the post diversification era.  From the year 2008-09 farmers adopted crop diversification and they started using chemical fertilizers. In these villages, the farmers use different types of fertilizers for different purposes. They use POTAS for the growth in the size of the fruits and DAP for the growth of the plants and URIA to maintain a good soil health. Different ratio and different quantities of chemical fertilizers were used by the farmers in the study areas but they still continued with the plough man and traditional bullock cart technology. So it can be said that this was the first step towards their cultural diversity.
 
Coming to agricultural implements, the farmers in the study area used plough to till the land with a combination of 2 bullocks and a man. With this combine effort a man days created, what we have stated in Tab. \ref{Tab:AgriculturePrctices}.
From the year 2013-14 and onwards,
 they have also adopted modern implements like tractors for their cultivation purposes except in village 5. In that village they trusted on the labour intensive techniques. So we can say gradually they have adopted the modern techniques and they have changed their cultural beliefs.

\begin{table}[h!]
\centering
\caption{cost of agricultural inputs per acre (prices are in Rupees). cow-dung manure was used in 2004-05, while they switched to chemical fertilizers onwards.}
\begin{tabulary}{\textwidth}{L| L| L| L| L| L| L}  
   \hline
Year &Use of agricultural inputs &Village 1 &Village 2 &Village 3 &Village 4 &Village 5 \\
   \hline
 2004-05 &Price of fertilizer (cow-dung manure) &1200 &1400 &1000 &1400 &800\\
     \cline{2-7}
~ &Agricultural implements (price for single man days)&200 &170 &150 &150 &120\\
\hline
\hline
2008-09 &Price of fertilizer (chemical and cow-dung manure)  &3900 &2400 &6200 &2200 &1200\\
     \cline{2-7}
     ~ &Agricultural implements (price for man days) &500 &440 &300 &400 &300\\
\hline
\hline
2013-14 &Price of fertilizer (chemical) &4700 &3800 &5600 &2600 &2700\\
     \cline{2-7}
~&Agricultural implements (price to hire tractor per two hours) &950 &1000 &950 &1200 &450 (man days)\\
     \hline
     \hline
2018-19 & Price of fertilizer (chemical) &3200 &5800 &7300 &4300 &2900\\
     \cline{2-7}
~&Agricultural implements (price to hire tractor per two hours) &1400 &1200 &1700 &1600 &1200\\
 \end{tabulary}
\label{Tab:AgriculturePrctices}
\end{table}


\subsection{ Social practices:}
In a tribal village each and every person are dependent on each other for their social and economic life. Here in this Tab. \ref{Tab:SocialPrctices} we can see they have a social set up and they exchange their man days for their work as building maintenance, crop cutting, messengers, cooking fuel (wood) in the year 2004-05. Here we have mentioned the calculated average payment as stated by the respondents.

We all acquainted with the concept of barter system, they use it in practice \cite{barterSystem}. They help each other in their needs so that they don't exchange money for work they exchange labour. This we have explained as Man days in the Tab. \ref{Tab:SocialPrctices}.

Messenger is a part of their social set up and acts like a postman. He receives and delivers the messages from one village to another from one family to their relatives. For this work he was paid a minimal monetary benefits, or may be in day to day things. In addition to this he has a duty to inform the villagers about their village meetings by striking the drums throughout the village. With the effect of modernization now the villagers are using cell phones as an easy way of communication. Even the messengers don't like to pursue their tradition job practice in the modern days. We have given an estimate of this per person or per households respectively in ab. \ref{Tab:SocialPrctices}.

\begin{table}[h!]
\centering
\caption{expenditure for of social set up (payment per day, prices are in Rupees). In 2004-05, we have stated the price of the work per day but they exchanging their labour days for each other. Same continued for 2008-08 except for Maintenance of Mud hut. This practice discontinued from the year 2013-14 and onwards.}
\begin{tabulary}{\textwidth}{L| L| L| L| L| L| L}  
   \hline
Year &Expenditure heads  &Village 1 &Village 2 &Village 3 &Village 4 &Village 5 \\
   \hline
2004-05 &Price for maintenance of Mud hut (per person) &120 &150 &100 &120 &100\\
     \cline{2-7}
~ &Price for crop cutting (per person) &120 &150 &100 &120 &100\\
     \cline{2-7}
~ &Messengers fees  (per person) &40 &30 &40 &20 &20\\
     \cline{2-7}
~ &fuel (wood) expenditure (per household) &15 &8 &10 &8 &5\\
\hline
\hline
2008-09 &Price for maintenance of Mud hut (per person) &150 &150 &140 &120 &120\\
     \cline{2-7}
~ &Price for crop cutting (per person) &120 &150 &100 &120 &100\\
     \cline{2-7}
~ &Messengers fees  (per person) &50 &30 &60 &50 &40\\
     \cline{2-7}
~ &fuel (wood) expenditure (per household) &20 &15 &15 &15 &10\\     
\hline
\hline
2013-14 & Price for maintenance and Construction of Pucca house (per person)  &240 &200 &240 &220 &200\\
     \cline{2-7}
~ &Price for crop cutting (per person) &180 &170 &180 &150 &150\\
     \cline{2-7}
~ &cell phones recharges (per household) &10 &5 &10 &10 &2\\
     \cline{2-7}
~ &fuel (Gas) expenditure (per household) &20 &20 &15 &15 &12\\
     \hline
     \hline
2018-19 & Price for maintenance and Construction of Pucca house (per person) &300 &300 &300 &250 &250\\
     \cline{2-7}
~ &Price for crop cutting (per person) &220 &200 &180 &200 &180\\
     \cline{2-7}
~ &cell phones recharges (per household)  &15 &10 &15 &15 &10\\
     \cline{2-7}
~  &fuel (Gas) expenditure (per household) &20 &20 &15 &15 &15\\
\end{tabulary}
\label{Tab:SocialPrctices}
\end{table}

In Tab. \ref{Tab:SocialPrctices} we can see, the changes can be found partially regarding the maintenance of the building in the study area and of crop cutting, messengers, cooking fuel they exchange their man days but the price of the man days were increased.
We can experience a complete change in social set up after 2013-14. New technology were adopted the tribal people started adopting the use of cell phones and switched from the use of cooking fuel of wood to the use of cooking gas.

\subsection{Traditional and cultural practices:}
It appears one in-separable part of tribal life's traditional and cultural practices is, Alcohol and tobacco.  Alcohol made by fermentation process (Mahuli from Mahua flowers,  ), or taking juice from wing of tree and later intoxicating it by traditional means (Salapa from Salap tree, date palm), rice beers (Pendum, Handia ) and from distillation as well (Tamarind, Mango and molasses). One peculiar thing is that, other liquor may be brought with barter system, but Salapa is available on cash payments and this is considered to be a Royal drink in tribal society \cite{buti_book}.  Tobacco is used in cigar form, where they roll tobacco leaves or small pieces of tobacco leaves used with Calcium Carbonate for chewing. One can find from Tab. \ref{Tab:TraditionalPrctices} that, alcohol and tobacco amounts almost double, from rest of items used related to traditional and cultural practices. 


They observe different kind of festivals, mostly inclined to cultivation (Nuakhai), for hunting (Chaiti Parab), worshiping mother nature (Ganta Parab,Magha Parab) or recreation (Pus-Parab). These are mostly associated with offering to mother nature, either food, wine or animal sacrifice. Offerings to god are distributed among all villagers. Their expenditure related to buying of animals, sending gifts to fellow villagers or in wine/tobacco.

Ceremonies and rituals are the important part of tribal life \cite{tribeRitual}. The ceremonies are of two types. In the family ceremonies are observed during birth, marriage and during funerals. The relatives gathered to name the baby on the 21st day. Prayer to the ancestors is made to seek the blessings for the newly born. Also they have the believe that after life there is a ritual to perform by making rice and curry etc with the guidance of Priest(Sisa). Here in Tab. \ref{Tab:TraditionalPrctices} it is shown that the average expenditure per ceremony.
\begin{table}[h!]
\centering
\caption{expenditure cost for different traditions per ceremony (prices are in Rupees). In the year 2004-05, they pay in terms of kinds for birth/life and marriage. In the year 2008-09 this was for life only. This practice discontinued from the year 2013-14 and onwards.}
\begin{tabulary}{\textwidth}{L| L| L| L| L| L| L}  
   \hline
Year &Expenditure heads  &Village 1 &Village 2 &Village 3 &Village 4 &Village 5 \\
   \hline
2004-05 &Birth of a child  &60 &50 &100 &30 &40\\
     \cline{2-7}
~ &Funeral processes  &30 &20 &40 &30 &30\\
     \cline{2-7}
~ &Marriage  ceremonies  &60 &50 &100 &30 &40\\
     \cline{2-7}
~ &Festival expenditure  &180 &230 &220 &200 &100\\
     \cline{2-7}
~ &  Tobacco and Wine  &600 &900 &600 &600 &300\\
\hline
\hline
2008-09 &Birth of a child  &100 &50 &100 &50 &50\\
     \cline{2-7}
~ &Funeral processes  &30 &20 &40 &30 &30\\
     \cline{2-7}
~ &Marriage  ceremonies   &100 &80 &150 &80 &50\\
     \cline{2-7}
~ &Festival expenditure  &200 &250 &220 &200 &100\\
     \cline{2-7}
~ &  Tobacco and Wine   &1000 &1200 &1500 &800 &600\\
  \hline
\hline
2013-14 &Birth of a child  &120 &80 &100 &100 &70\\
     \cline{2-7}
~ &Funeral processes  &50 &50 &60 &50 &30\\
     \cline{2-7}
~ &Marriage  ceremonies   &200 &150 &250 &100 &100\\
     \cline{2-7}
~ &Festival expenditure  &500 &700 &600 &200 &800\\
     \cline{2-7}
~ &  Tobacco and Wine   &2000 &1500 &1500 &1200 &1000\\
  \hline
\hline
2018-19 &Birth of a child  &200 &250 &100 &100 &100\\
     \cline{2-7}
~ &Funeral processes  &50 &50 &60 &50 &30\\
     \cline{2-7}
~ &Marriage  ceremonies   &200 &150 &250 &100 &100\\
     \cline{2-7}
~ &Festival expenditure  &850 &1000 &2000 &700 &800\\
     \cline{2-7}
~ &  Tobacco and Wine   &2000 &1500 &1500 &1200 &1000\\
\end{tabulary}
\label{Tab:TraditionalPrctices}
\end{table}

It can be seen that there is a change occurred in 2008-09 that people changed their traditional way of sharing gifts during the all the aspects except the case of funeral. But subsequently they made a complete change and started spending more and more during their festivals and ceremonies during the years 2013-14 and 2018-19. 


\subsection{Approach towards to modern life:}
Indigenous knowledge of night sky and weather forecast was already available with tribal people. They also used dialects, related to nearby prevailing languages. Government through the department of Tribal and Rural Welfare have been trying its best for the educational development of the tribals \cite{GovtAttempts}. Various schooling institutes (Aadasrh Vidyalayas, Ashram Schools, Sevashram) and Training centers have opened by the government for educating the tribal people. Stipends are granted to tribal students for higher education. They get all the fees reimbursed by the govt. so the expenditure mentioned in Tab. \ref{Tab:ImpactCDI} is the allied expanses such as purchase of books and notes, uniform, pocket money, etc.

\begin{table}[h!]
\centering
\caption{cost of average expenditure per households (prices are in Rupees), after a complete approach to the modern techniques, i.e. 2013-14 and 2018-19.}
\begin{tabulary}{\textwidth}{L| L| L| L| L| L| L}  
   \hline
Year &Expenditure heads  &Village 1 &Village 2 &Village 3 &Village 4 &Village 5 \\
   \hline
2013-14 &New clothing during festivals  &2600 &1600 &4250 &2200 &1500\\
     \cline{2-7}
~ &Medical treatments  &628 &479 &782 &1142 &427\\
     \cline{2-7}
~ &Education  &2643 &416.4 &247.02 &1157 &321.41\\
     \hline
     \hline
 2018-19 &New clothing during festivals   &4800 &3600 &6400 &2700 &2400\\
     \cline{2-7}
~ &Medical treatments   &1378.64 &938.52 &1522.39 &1800 &625\\
     \cline{2-7}
~ &Education  &4660.19 &737.7 &895.52 &3480 &561.76\\
\end{tabulary}
\label{Tab:ImpactCDI}
\end{table}

After the growth in the income of the farmers now they are maintaining a better standard of living and spending a sizable amount for the purchase of clothes on some of the important festivals they observe. Buying of new clothes on occasion of major festivals (Chaiti Parab, Magh Pus Parab and Ganta Parab) is a recent trend, towards their approach to modern life. 
This practice is became common due to the increase in income though crop diversification.

Among the most primitive tribe of Odisha, there is the belief that disease is caused by hostile spirits, the ghosts of the dead or due to the violation of some taboo. Indigenous methods of treatment of the diseases among the tribal can be divided into two categories namely 1. Magical cure and 2. Medicinal cure. In case of epidemics like small pox, cholera or cattle disease they believe that it is caused by the evil influence of the Duma (Ghost). The family has to celebrate worship to village goddess (Thakurani). 
They use to go to their traditional doctor (Disan) but now-a-days the belief has been changed and now they use to come to purchase the medicines as prescribed by the certified physicians by  govt. This change can be understood from Tab. \ref{Tab:ImpactCDI}.

\section{Results and discussion}
\subsection{Simple linear regression formalism}
A simple linear regression (SLR) may be written as:
\begin{align}
Y &= b_0+b_1X \\
where \
b_0 &= \bar{Y} - b_1 \bar{X}, \\
and \
b_1 &= \frac{N\sum XY - \sum X \sum Y}{N\sum X^2 - (\sum X)^2}
\end{align}

We have used CDI from our earlier work Ref. \cite{cdi_mypaper} as independent variable (X) and different factors of socio-economic life taken as dependent variable (Y), such as Agricultural practices, Social practices, Traditional and cultural practice and modern life practices. We have added individual sub-factors contributing to each of these dependent variables. Except for Tab. \ref{Tab:SocialPrctices}, where messenger fee/cell phones or fuel have not added to the sum. As the unit for measurement (per person or per household) is not consistent with other entries of Tab. \ref{Tab:SocialPrctices}. SLR coefficients are presented in Tab. \ref{Tab:RegressionCoefficientsVillageWise}.

\begin{table}[h!]
\centering
\caption{Regression coefficients}
\begin{tabulary}{\textwidth}{L| L| L| L| L| L| L}  
 \hline
 Expenditure heads &Regression coeff.  &Village 1 &Village 2 &Village 3 &Village 4 &Village 5 \\
 \hline
 Agricultural practices & b0 & 1957.86 &1738.68 &2187.97 &842.231 &1159.66\\
 \cline{2-7}
 ~ &b1 & 16041.6 &6256.54 &12457.9 &9784.28 &2738.27 \\
 \hline
 \hline
 Social practices & b0 &168.798  &199.203 &199.757 &125.503 &170.854\\
 \cline{2-7}
 ~ &b1 &1552.76 &394.525 &346.719 &682.633 &244.519 \\
  \hline
 \hline
Traditional and Cultural practices & b0 &783.394 &977.507 &1235.07 &468.62 &589.538\\
 \cline{2-7}
 ~ &b1 &11819.8 &2823.75 &3368.13 &3559.82 &1713.18 \\
   \hline
 \hline
Modern life expenditures & b0 &-71.998  &967.125 &2415.84 &185.212 &403.542\\
 \cline{2-7}
 ~ &b1 &43737.5 &2443.81 &2050.28 &11727.3 &2699.9 \\
\end{tabulary}
\label{Tab:RegressionCoefficientsVillageWise}
\end{table}

\begin{figure}
\centering
\includegraphics[scale=0.5]{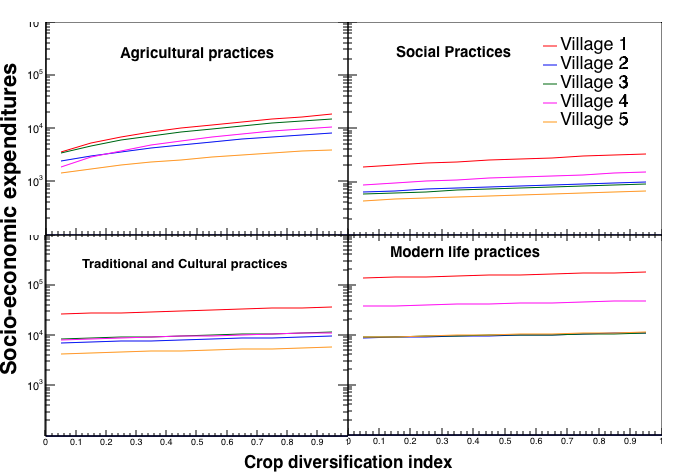}
\caption{(Color online) CDI for different villages, calculated from SLR equation for all five villages}
\label{fig:CDIVsSLREq} 
\end{figure}

To vsualize this effect, we have plotted CDI with different dependent variables in Fig. \ref{fig:CDIVsSLREq}. We can see there is a positive impact of crop diversification on the socio-economic life of the farmers observed in the study area. We have measured the impact of crop diversification on the individual variables. The impact of crop diversification on modern life practices is the highest in case of village-1, village-2 and village-4. In case of village 3 and village-5 the highest impact found on agricultural practices. We can add here that there is comparatively less impact found on the expenditures of social practices and traditional and cultural practices. The reason behind that village-3 and village-5 go on diversifying their cropping pattern out of which the expenditure on agricultural practices has been increasing. For village-1, village-2 and village-4 they are focusing more on Education, Health and; clothing as we can see in the table-1, the literacy rate of these 3 villages is comparatively higher.

\subsection{Coefficient of Correlation}
We have calculated Coefficient of Correlation by Karl Pearson's method in Tab. \ref{Tab:CoefficientofCorrelation}. We observe about 10\% negative correlation for village 3 for CDI, for Modern life expenditures. Similarly social practices have minimal degree of correlation between for village 2 and 3 for CDI. For village 1, we see almost a perfect correlation between CDI and Traditional and Cultural practices/Modern life expenditures. Most of variables show a moderate degrees of correlation for Village 4 and 5, traditional and cultural practices being most prominent.

\begin{table}[h!]
\centering
\caption{Coefficient of Correlation by Karl Pearson's method}
\begin{tabular}{ c|c|c|c|c|c}  
~  &Village 1 &Village 2 &Village 3 &Village 4 &Village 5 \\
\hline
  Agricultural practices &0.683943 &0.27873 &0.57871 &0.639337 &0.686499\\
\hline
  Social practices   &0.966471 &0.0246893 &0.010771 &0.483779 &0.669521\\
  \hline
Traditional and Cultural practices  &0.980217 &0.273141 &0.233766 &0.635929 &0.842781\\
  \hline
Modern life expenditures   &0.948318 &0.0523699 &-0.104818 &0.475428 &0.664583\\
\end{tabular}
\label{Tab:CoefficientofCorrelation}
\end{table}

\section{Conclusion and policy recommendations} 
\label{Summary}
In this study, we have investigated impact of crop diversification on various socio-economic aspects of tribal people from eastern ghats of India.
In this paper we found there is a positive and proportionate impact of crop diversification on the expenditure habits of the tribal farmers in the study area. The farmers are especially spending on the agricultural practices and modern life practices which is the need of the day. 

Govt. or other development agencies should make targets to make their culture sustainable, while their economic developments grows. Policies can also be made to provide funds to the maintenance of cattle so that the farmers will not sell them and again the cow dung will be available in plenty and the pace of using bio-fertiliser can be accelerated.




\end{document}